# Efficient proton acceleration from laser-driven cryogenic hydrogen target of various shapes

A.Sharma[1,a)], A. Huebl[2,3], A.Andreev[1,4]

1. ELI-ALPS, Szeged, Hungary.
2. HZDR, Dresden, Germany.
3. TU Dresden, Dresden, Germany.
4. Max-Born Institute, Berlin, Germany.

We theoretically investigate high energy – collimated proton beam with three dimensional particle-in-cell simulations of ultra-short petawatt laser interaction with cryogenic hydrogen target of various shapes. Here we show that under appropriate conditions between the laser and target parameters, the protons are accelerated to high energies mainly due to collisionless shock acceleration mechanism combined with TNSA. The dependence of the protonic energy on the laser field, target shape and thickness is reported. It is demonstrated that the irradiation of intense laser (20fs-2PW) with cryogenic hydrogen target at optimal thickness allows the efficient generation of high energy proton beam (>100 MeV) of small divergence. Our results also indicate that diffracted laser field strongly affects the collimation of electrons/ions as it passes beside the mass limited target. This approach predicts a possible pathway to control laser driven ion sources.

a) Email: ashutosh.sharma@eli-alps.hu

**INTRODUCTION**

Petawatt laser-driven ion acceleration has attracted great deal of attention due to diverse prospects in the field of inertial confinement fusion, cancer therapy and particle accelerators [1-4]. Immense interest has been paid to laser-driven ion acceleration, which potentially offers a compact, cost-effective alternative to conventional sources for scientific, technological, and health-care applications [5-7]. Most experimental research, so far, has focused on the target normal sheath acceleration (TNSA) mechanism [8-10]. TNSA ion beams typically have a broad energy spectrum, modest conversion efficiency at high energies and, large divergence. Different other ion acceleration mechanisms e.g., radiation pressure acceleration (RPA) [11-19], shock wave acceleration (SWA) [20-22] and laser breakout after-burner [23] have recently drawn a substantial amount of experimental and theoretical attention[4] due to the predicted superior scaling in terms of ion energy and laser-ion conversion efficiency. Gas targets, also open the way for novel ion acceleration e.g., magnetic vortex acceleration (MVA) [24-25], which is suitable for high repetition-rate operation. To achieve high energy proton beams via the MVA mechanism, a tightly focussed laser beam and near critical density plasma with sharp density gradient is required – a non-trivial technical challenge. Most of the potential applications require high energy, high quality proton and ion beams with high collimation, high particle flux and monoenergetic features. Consequently, the beam quality enhancement is highly important and thus, there have been numerous experimental and theoretical studies [4] working to this goal [1-7].

Mass-limited targets (MLT) [26-28] have also attracted attention due to expected enhancement in efficiency and maximum cut-off ion energy, when compared to plane target of similar dimensions. Andreev *et al*. [26] have studied the laser driven ion acceleration from MLT droplet via 2D3V particle-in-cell (PIC) simulations and reported significant enhancement in ion energy under the optimum condition of laser beam and target diameter. The experimental, PIC simulation and analytical model presented by Sokollik *et al.* [27] explained the limitation of using spherical MLT and reported low energy ion in their experiments. These limitations can overcome by using high contrast laser pulses with MLT. Their investigation also suggested the optimum condition to enhance the ion energy by considering comparable spherical diameter and laser spot size. In the 2D PIC simulation study, Zheng et al. [28] investigated the generation of fast electrons and protons using weakly relativistic laser pulses. They reported proton acceleration by the electrostatic field induced by the hot electron jet due to resonance absorption and isotropic proton acceleration by ambipolar expansion. Psikal et al. [29]

investigated theoretically via PIC simulation the ion acceleration by ultrashort intense femtosecond laser pulses in small targets of uniform chemical composition of two ion species (protons and carbon ions); where dips and peaks are observed in proton energy spectra due to mutual interaction between two ion species. Lucchio *et al*. [30] investigated MeV ions from nano-droplets target driven by a few cycle laser pulse using 2D PIC simulations. The use of cryogenic hydrogen targets reduces the accelerated species to only protons and additionally produces a higher accelerating field due to MLT.

The crucial feature of MLT is the limited target size, which leads to a confinement and recirculation of plasma electrons resulting in an additional interaction with the laser pulse, which changes the energy distribution function and enhances the ion energy. When the dimension of MLT and the laser spot size are comparable and irradiated by a short and ultra-intense laser field, the ions can be accelerated together with the electrons by the radiation pressure dominated acceleration (RPDA) [18] mechanism up to an energy substantially higher than the energy achievable in the case of flat target [31].

Nowadays, laser technologies capable of producing high proton energies may enable further investigations into the new regime of ion acceleration using a cryogenic hydrogen target without debris. The availability of a cryogenic hydrogen target [32], using technology developed at SLAC may provide a pure, continuous, mass-limited target that will not be subject to problems like energy spread or contamination. A custom made cryogenic target mount cooled by a cold head down to temperatures as low as 8 K can be used for the production of hydrogen targets The various target geometry can be determined by the geometry of target mount and growing chamber however thickness can be reduced by controlling the heating. Recent experiments by Propp [32] at TITAN with a pure liquid near-critical density jet, where a 527 nm split beam, frequency-doubled TITAN laser produced a pure proton beam with monoenergetic features. The data from the cryogenic jets was limited due to the heating of jet and orifice damage. Recent experimental [33] investigations reported the 20 MeV peak proton energy with $10^9$ particles per MeV per staradian, while employing a continuous cryogenic hydrogen jet with 150 TW ultrashort laser pulse Draco.

We investigate with 3D PIC simulation the efficient approach of proton acceleration driven by ultra-short and relativistic-intense laser field and illustrate the possibility to comparable advanced laser facilities e.g. ELI-ALPS' High Field laser [34]. We optimised the MLT targets of various shapes, which may play a crucial role in enhancing the proton beam properties in comparison to use of foil and gas targets. The simulation study reveals involved acceleration mechanism, optimum conditions and scaling for high energy proton beam and high number of protons. We also show the influence of laser polarisation on proton beam characteristics, as a function of proton energy. It is also delineated that diffracted laser field beside the MLT target can shape the proton beam to make it appropriate for medical applications.

**THEORETICAL FRAMEWORK AND SIMULATION METHOD**

We start by describing the interaction of a linearly polarized laser with the plasma medium, which is considered the pre-ionized cryogenic hydrogen target of different geometry. The irradiation of the front surface of the target causes the electrons in the skin depth from the front surface to be accelerated by the ponderomotive force [26, 35]. To investigate the proton acceleration at ultra-short laser interaction with cryogenic hydrogen target we have performed 3D PIC simulation with the code PIConGPU [36]. Geometrical factors of a MLT should be displayed in higher dimensionality in order to accurately explore the acceleration mechanism. An ideal laser pulse (800 nm Ti: Sapphire laser system) is considered; which is Gaussian in space and perfect contrast in time. The beam diameter is 2.5 µm and laser pulse duration is 20 fs. The linearly polarized laser is focused at the front side of target along the laser propagation direction (y-axis) and the peak laser intensity is ~ $10^{22}$ W/cm$^2$. The proposed cryogenic hydrogen plasma targets of different geometry (planar, cylindrical and spherical) and size (diameter 1 – 5 µm) are considered however the plasma density is kept close to $n_i = n_e = 6.96 \times 10^{22}$ cm$^{-3}$ (~ 40$n_c$). A simulation box of dimension 10×10×10 µm$^3$ is considered corresponding to the grid size 1024×1024×1024 with cell size of 10 nm and the time step is 16.7 asec. The number of particles per cell is 2 in each direction. Periodic boundary conditions are used in the simulations in transverse direction (along x and z-axis) and absorbing along the laser propagation direction (y-axis). The open/absorbing boundary condition is used along the laser propagation direction to reduce the computational costs. The peak density in simulation increases up to 2-3 times the initial plasma density so the grid size and time step are chosen carefully to resolve the electron dynamics within the relativistic collisionless skin depth (~$\gamma^{1/2} c / \omega_{pe}$), where ω$_{pe}$ is an electron plasma frequency. The linearly polarized laser is opted in this study and further comparisons are made with the circularly polarized laser differentiate the acceleration mechanism. To delineate the dependence of maximum proton energy on laser field the laser power is varied while keeping the beam diameter constant to maintain the optimized condition for maximum ion acceleration. The targets are considered ideal (no pre-plasma expansion effects).

# SIMULATION RESULTS

The simulation results of the interaction of intense laser field $E_L$ ($a_L = eE_L/m_e c\omega = 71$) with the cylinder (by supposing the length of cylinder $L_t = 9$ µm much longer than the diameter of cylinder $D_t = 2.5$ µm) are shown in Fig.1. The laser - plasma parameters used in the simulation are determined above.

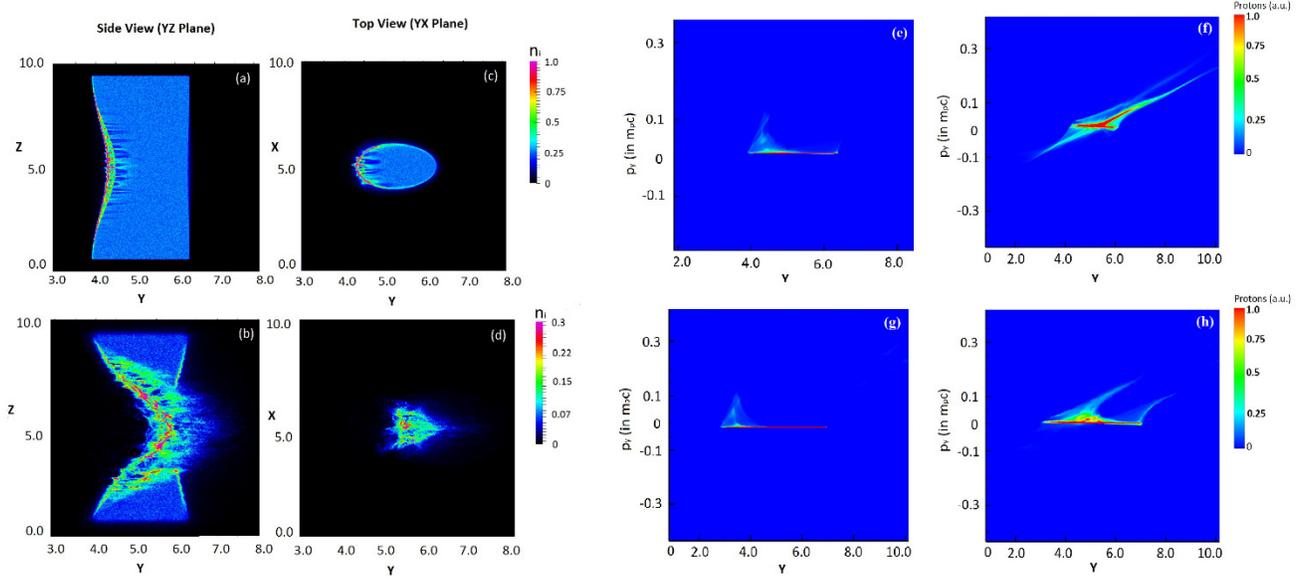

**FIG. 1.** (a-d) Ion density distribution for a long cylindrical target ($L_t = 9$ µm and $D_t = 2.5$ µm) at time instant ($\omega_{pi} t = 18$) (corresponding to top row) when the peak laser field interacts with the target while the bottom row shows the ion density distribution at time instant ($\omega_{pi} t = 30$) when ions achieve peak energy. Colour bar shows the variation in ion density which is expressed in units of $n_{cr}$ where $n_{cr} = \gamma n_c$ and $\gamma = \sqrt{(1+a_L^2/2)}$, $\omega_{pi}$ is an ion plasma frequency. (e-h) Evolution of the ion phase space at $\omega_{pi} t = 18$ (e, g), 30 (f, h) for cylindrical target of diameter 2.5 µm (e-f) and 4 µm (g-h). $\omega_{pi}$ is an ion plasma frequency.

As the linearly polarized laser starts interacting with the cylindrical target, the stable component of the ponderomotive force drives electrons forward and the high-frequency oscillation keeps heating electrons. The ion density distribution shown in Figs. 1a & b corresponds to the initial hole boring (HB) stage [35] due to the radiation pressure of the laser field. The target ($n_e = 40 n_c$) is relativistically under-dense ($n_e/\gamma n_c = 0.56$) for the ultraintense laser field ($a_L = 71$) and the interaction of linearly polarized laser with the significant volume of MLT target allows the efficient heating of the electrons due to the oscillating component of the laser field. Figs. 1 c & d shows the ion density distribution at time instant ($\omega_{pi} t = 30$) when ions achieve maximum energy. These simulation results shows that the critical density surface of cylindrical target is pushed forward by the laser radiation pressure with speed 0.2c and the ions reflecting from this shock potential will reach the rear surface of target ~0.4c. The ions with speed 0.4c may reach the target rear side (by travelling the distance ~ 2.4 micron) within the laser pulse duration.

Figs. 1(e-h) reveals the phase-space evolution of optimal acceleration by laser pulse. Its show a temporal evolution of the ions in momentum space in which the protons, accelerated by collective electrons from the front surface, are faster than the thermal accelerated electrons at the rear surface (Fig. 1 e-h), resulting in higher proton energy. An electrostatic shock wave is generated from the target front surface and propagates through the target. The shock, generated at the front surface with a velocity close to the HB velocity [35], is consistent with the assumption that shock waves are driven by the piston action of radiation pressure. Thus, the laser field and charge separation field propagate further in the plasma and combine with the TNSA field at the rear side of the target and this superposition amplifies the accelerating field at the rear side of the target, which results in higher proton energy (Fig. 1f). For comparison, the momentum distributions of protons in Fig. 1 g-h show non-optimum conditions where protons from front surface are not able to reach the rear surface to achieve the maximum proton acceleration (Fig. 1g-h).

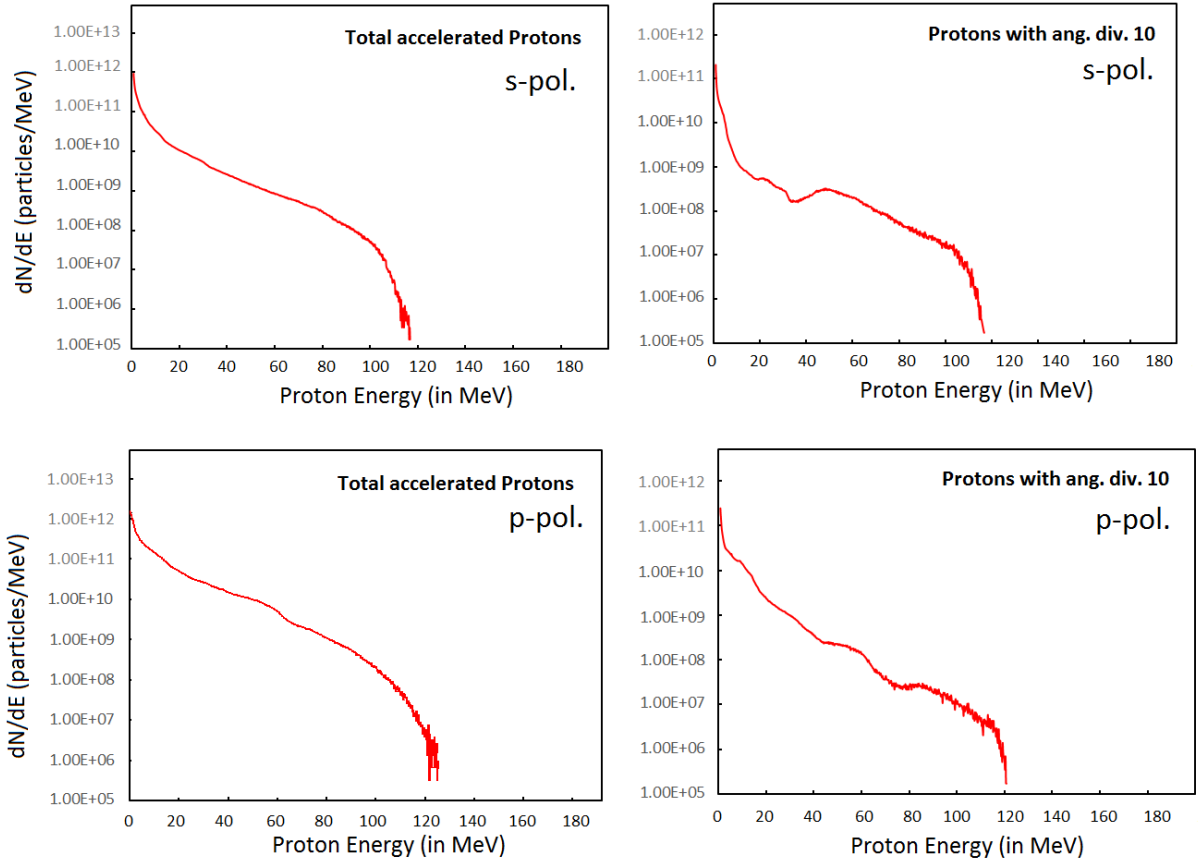

**FIG. 2.** Proton energy distribution at time instant $\omega_{pi} t = 30$ $a_L$=71, $D_L$= 5 µm, $n_e$ = 6.96x10$^{22}$ cm$^{-3}$ and size of plasma target: $L_t$= 9 µm, $D_t$= 2.5 µm. The laser electric field for s-polarised case is polarised perpendicular to the incident plan (XZ) and in case of p-polarisation the laser electric field is polarised parallel to the incidence plane (XZ) where the cylinder axis is considered along the z-axis.

Figure 2 shows the proton energy spectrum (top row corresponds to s-polarisation and bottom row for p-polarisation) where the incident laser intensity on the target is $1.1 \times 10^{22}$ W/cm$^2$ for long cylindrical target (target length = 9 µm) at target diameter of 2.5 µm. Separate simulations are made for s-polarisation and p-polarisation to investigate the effect of polarisation on proton energy distribution. The energy distribution of total accelerated protons and protons at divergence angle of $\theta < 10^o$ (right side plot) was measured from simulation results, where $\theta$ is defined as $\tan(\theta) = p_\perp / p_\parallel$ with $p_\perp$ and $p_\parallel$ are the transverse and parallel components of the proton momenta, respectively. It can be seen from the proton energy distribution (even at $\theta < 10^o$) that there are higher number of protons ($>10^9$) at energies $E > 20$ MeV, which may be sufficient for many applications [37-43]. It is clearly evident from the energy distribution and spatial density distribution of protons that more protons are in the high energy range for the s-polarized laser field but the peak proton energy is slightly higher for the case of p-polarization.

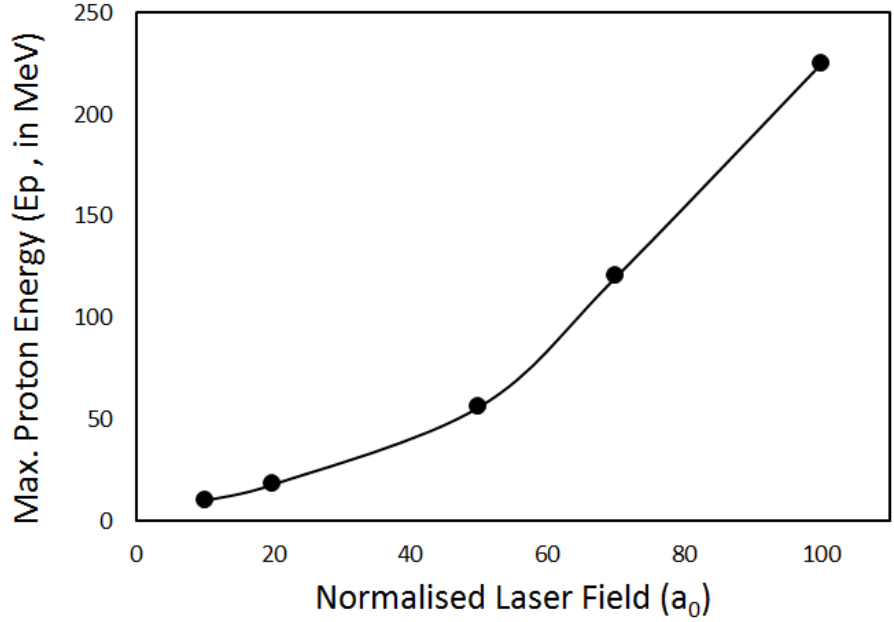

**FIG. 3**. The proton energy dependence on normalized laser field where protons are accelerated by linearly polarized 20fs – 2PW laser pulse incident on the cylindrical target (length - 9 µm) of optimized thickness 2.5 µm.

In order to use this acceleration regime in proton radiotherapy, where ~ 200 - 300 MeV proton energy [7] is needed, the scaling of the dependence of peak proton energy $E_p$ with the normalized laser field (by varying the laser intensities) for the cylindrical target case (see Fig.3) of optimum thickness was investigated. In this case the protons propagating close to the propagation axis (~ at divergence angle of 10 degree) were considered of interest for practical purpose [1-7]. The proton cutoff energy is strongly dependent on laser field and the peak proton energy scales with the normalized laser field as $E_P \propto a_L^\kappa$ (where κ =1 when $a_L$ < 50 and κ >1 at $a_L$ > 50), as shown in Fig. 3. The laser field partially penetrates the target when the normalized laser field is close or higher than the target parameter $n_e / n_c = 40$ and hence transparency plays important role in enhancing the proton energy.

In order to emphasize the practical utilization of high-repetition high-power short laser pulses in this regime, the scaling of the peak proton energy and proton numbers with the target thickness for plane, cylindrical and spherical target is considered. The 2 PW laser pulse length is fixed at 20 fs and the initial parameters were $a_L$ = 71, laser beam spot size $D_L$ = 5.0 µm, $n_e$ = 6.96x10$^{22}$ cm$^{-3}$ and thickness of plane target, diameter of cylindrical and spherical plasma target $D_t$ was changed. The peak proton energy and proton numbers corresponds to the protons propagating at 10°.

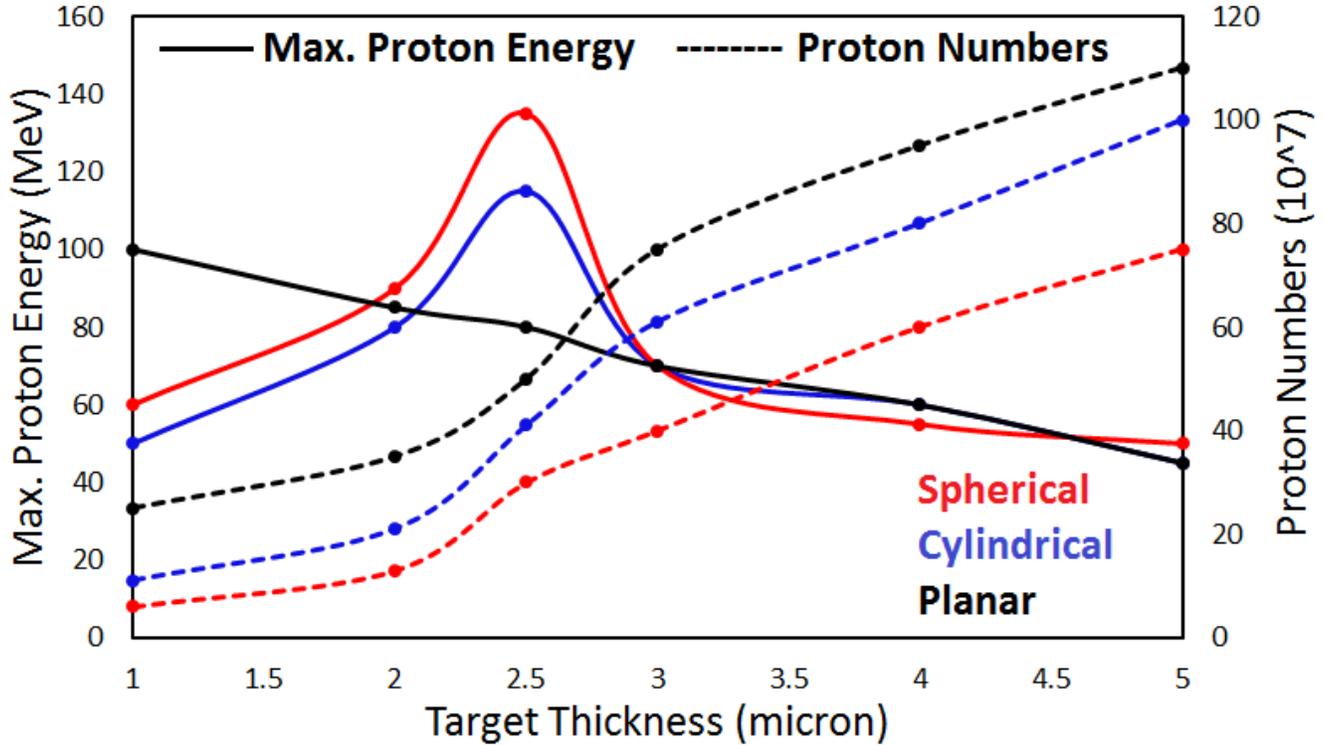

**FIG. 4.** Simulation results showing the dependence of peak proton energy and proton numbers on target thickness for different target geometry. In each case, the normalized laser field is, $a_L = 71$ generated by focusing the 2PW laser beam to 5 µm spot diameter. Red, blue and black curves correspond to spherical, cylindrical and plane target respectively whereas the solid curves represent to peak proton energy and dashed curves is shown for proton numbers. The simulation results shown are for protons propagating close at 10 degrees from the laser propagation axis.

The optimum condition of target thickness can be seen when the target diameter is 2.5 µm which is slightly less than half of laser pulse length; thus by considering the target thickness thin enough so that the ions accelerated at front surface due to SWA can reach the rear surface of target within the laser pulse duration. In such situation, ions at the front side accelerated due to SWA combine together with ions accelerated at rear surface due to TNSA which contributes in enhancing the peak proton energy.

Figure 4 shows the dependence of peak proton energy on target thickness for plane, cylindrical (length of cylinder - 9 µm) and spherical target. The simulation show that the optimum target size is nearly half of the laser pulse length for non-planar geometry and at optimum target diameter of 2.5 µm, the peak proton energy 135 MeV for spherical target shape and 120 MeV for cylindrical target is observed. In the 3D simulation, the decrease in peak proton energy for plane target which may be due to the dynamics associated with the laser-plasma interaction in front of the target and the formation of space charge field at the rear side of the target. These dynamics are different than the 1D scenario. In addition to the ion energy, it is also important to consider the total number of accelerated protons to the relevant energy ranges.

To continue the influence of target geometry on ion acceleration, the interaction of the laser field ($a_L$ = 71, $D_L$ = 5.0 µm) with spherical hydrogen plasma of solid density ($n_i = n_e$ = 6.96x10$^{22}$ cm$^{-3}$) and diameter $D_t$ = 2.5 µm is shown with identical laser-plasma parameters as in cylindrical target study.

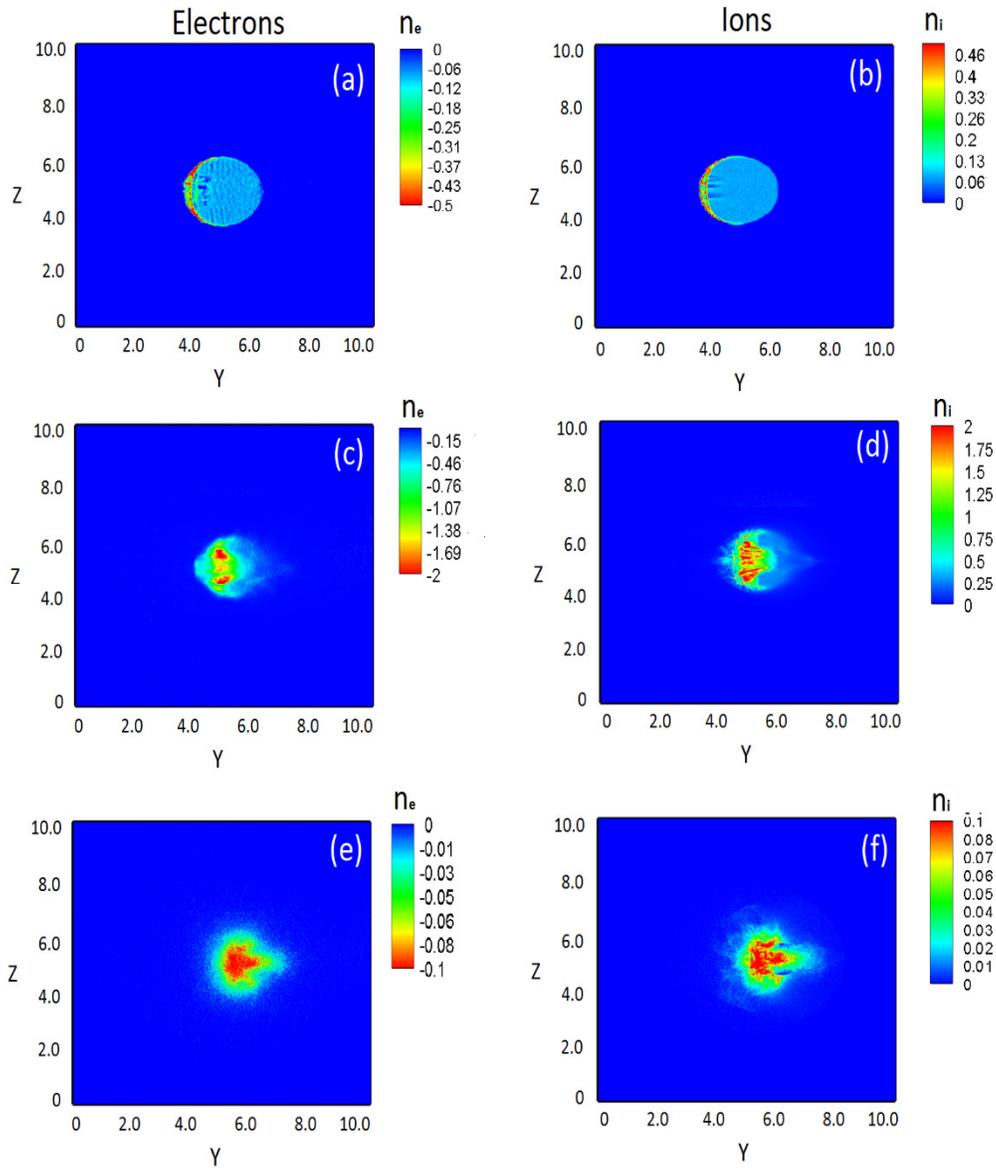

**FIG. 5.** The evolution of plasma density distribution in YZ plane for a spherical target at time instant

$\omega_{pi}t$ =18 (a-b) when the peak laser field interacts with the target, (c-d) at $\omega_{pi}t$ =24 and (e-f) when protons achieved peak energy ($\omega_{pi}t$ =30). The color bar shows the variation in electron/ion density with units of $n_{cr}$ where $n_{cr} = \gamma n_c$ and $\gamma = \sqrt{(1 + a_L^2/2)}$.

In order to understand the dynamics of plasma particles in spherical target driven under the influence of short and high power laser pulse, the electron and proton density distribution are delineated in Fig. 5. Fig. 5 shows the evolution of electron (a-c) and proton (b-d) density in YZ plane when the peak laser field is in regime of interaction with the plasma. Fig. 5 shows the evolution of electron (e) and proton (f) density in YZ plane at the time of maximum particle acceleration.

We observe from simulation results the collimation of ion beam at rear side of target, which can be attributed to the diffraction of ring shaped magnetic field of laser (explained later in Fig. 6). The signature of inhomogeneous ion density distribution (in Fig. 5 f) at the front side of target due to the fast evolving instability; at the rear side there is large collimated ion beam in center and inside there is small high density jets merging with the central ion beam. A possible explanation for these structures are Weibel-like instabilities [44] caused by counter streaming electron current – hot electrons which cannot overcome the electrostatic barrier return into the target and inside the target the cold electrons. This unstable regime leads to filaments at the target front surface. We observe the pronounced modulation in proton density in comparison of electron density distribution at time instant $\omega_{pi}t$ =30 (as shown in Fig. 5 e-f). The spatial density evolution of electrons (as in Fig. 5 a, c, e) and protons (as in Fig. 5 b, d, f) where the modulation in electron density (at $\omega_{pi}t$ =18) is mapped to the modulation in ion density modulation (at $\omega_{pi}t$ =24 and 30). Thus, the transfer of instability from the electron beam to proton beam can be explained by the faster thermalisation of the fast electrons due to the difference in mass. However, the larger mass of protons may result in the continuation of the instability after the electronic instability has already ended.

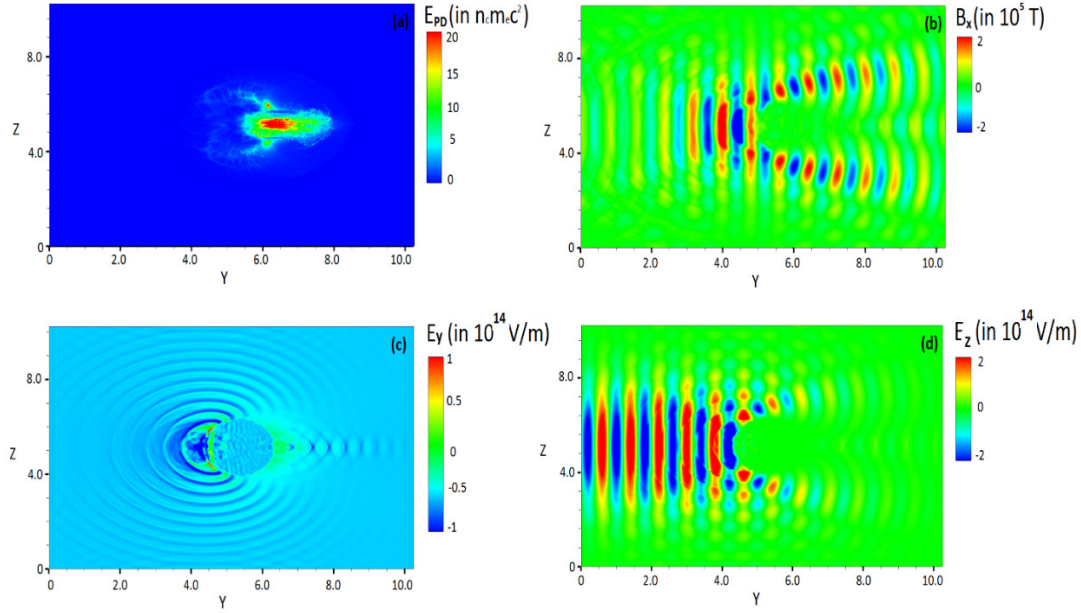

**FIG. 6.** Collimation of proton beam by the diffracted field of laser at time instant $\omega_{pi} t = 30$. (a) the energy density distribution of collimated proton beam where the color bar shows the variation in proton energy density ($E_{PD}$) normalised with $n_c m_e c^2$ (b) the magnetic field ($B_x$) distribution of diffracted laser field around the spherical target and color bar shows the variation in magnetic field which is in units of $10^5$ T (c) the distribution of longitudinal field ($E_y$) along the laser propagation direction and (d) the electric field ($E_z$) of laser at time instant $\omega_{pi} t = 18$. The color bar shows the electric field variation in units of $10^{14}$ V/m.

Fig. 6 (a-b) shows the collimation of proton beam due to the magnetic field component of diffracted laser field. At the rear of the target surface, the noninteracting diffracted electromagnetic field of laser and subsequently the radial ponderomotive force (RPF) provides the radial compression and confinement as well as directional stability of the ion beam. Initially the RPF generated by the non-diffracted Gaussian laser field distribution pushes the electrons radially outward. At a later stage of the interaction (time instant $\omega_{pi} t > 18$), the RPF generated by the diffracted laser field (non-Gaussian e.g. ring shaped intensity distribution with intensity at center is zero) pushes the electrons radially inward keeping the electron beam collimated at rear side of target. This can be concluded from the expression of density modulation due to the radial poderomotive force [45]: $n_e = 1 + k_p^2 \nabla^2 \gamma$, where $\gamma = \sqrt{(1 + a_L^2/2)}$, $k_p = \omega_p/c$ and $\omega_p$ is the plasma frequency.

The radial deflection of accelerating proton under the influence of helical magnetic field ($B_0$) can be written as $\delta = ev_z B_0 L / 2E_p$ where $v_z$ is the longitudinal velocity, L is the interaction length, $E_p$ is the proton energy. To maintain the divergence angle (divergence angle ~$\delta$ ), 10 degree of ~135 MeV protons for 2 μm, the estimated magnetic field amplitude is ~$7 \cdot 10^4$ T – similar to the magnitude shown in Fig. 6 (b) for magnetic field distribution.

The simulation results in Fig. 6 show the distribution of electric field ( (c) – longitudinal electric field, $E_y$ and (d) - electric field of laser $E_z$) when linearly polarised laser interacts with a spherical target is smaller than the laser beam. At time instant $\omega_{pi} t = 30$, the longitudinal component of electric field (Fig. 6 (c)) is of the same magnitude as the electric field of laser (Fig. 6 (d)) and the critical density surface shows an inward motion, i.e. ponderomotive force dominates over thermal pressure. However the motion of critical surface carries an imprint of laser absorption process validating the dominance of *jxB* absorption mechanism at relativistic laser intensity.

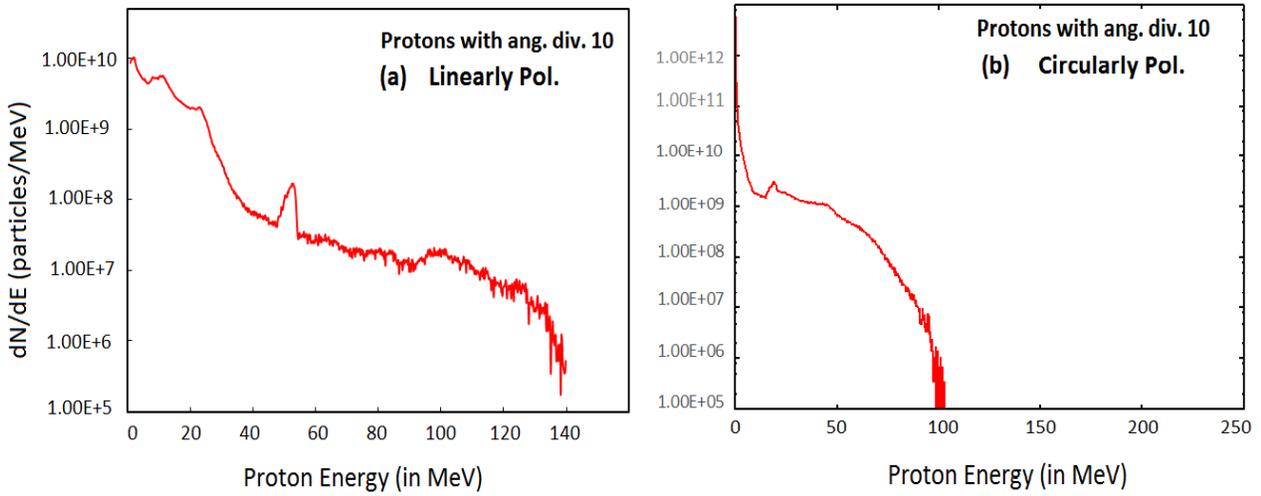

**FIG. 7.** a) Proton energy distribution at time instant $\omega_{pi} t = 30$ (a) linearly polarised laser light b) circularly polarised laser light. The laser-plasma parameters are as - $a_L$ =71 (Lin. Pol.) and 51 (Cir. Pol.), $D_L$= 5.0 μm, $n_e$ = 6.96x10$^{22}$ cm$^{-3}$ and, $D_t$ = 2.5 μm.

Figure 7 (a) shows the proton energy distribution for spherical target of diameter 2.5 micron interacting with the linearly polarised laser. Protons are considered in the above plots, which are accelerated close the propagation axis (along the laser propagation direction) at divergence angle of 10 degrees. Figure 7 (b) shows the proton energy distribution for circularly polarised laser at time instant when proton achieves maximum energy. In the case of circular polarisation, the ions are accelerated to lower peak energy in comparison to the linearly polarised case because of smaller force, as shown above. The past studies [4, 18, 46] of RPA with circularly polarised laser demonstrated the favourable enhancement of proton energy in comparison to linearly polarised laser while utilising the ultrathin targets of nanometre scale. We considered the cryogenic hydrogen targets of thickness on micrometre scale where an intense laser pulse bores a hole over the target skin depth and steepens the electron or directly the ions like a piston. This entails the ion shock within the target skin depth and as this shock propagates in the target, the ion bounce back at twice the shock velocity.

We focused in this proton acceleration research the target thickness (1µm - 5µm) of different geometries, i.e. plane, cylindrical and spherical . The minimum target thickness in this work is 1 µm, however, when thinner target (<1 µm) are used, the RPA scheme can apparently push the plane target quasi-monoenergetically with great efficiency but there is a limiting factor due to Rayleigh Taylor Instability [40] which limits the quality of ion beam. Transverse instabilities are less important in MLT due to their small targets [32] and thus should be realized for the stable ion acceleration in SWA regime. The simulation results presented in this article shows the signature of Weibel instability [44] in the proton beam emerging at front side of target while the instability is not influencing the high energy collimated proton beam emerging from the rear surface side.

**CONCLUSIONS**

In conclusion, we have investigated the significant enhancement in proton peak energy and proton flux via the 3D PIC simulation, utilizing the advance laser technology [34] and MLT target from the growing technology of cryogenic target development [32]. We have employed a high contrast, short and near-future intense laser field (20 fs - 2 PW) with ideal pre-plasma conditions which enabled the shock wave acceleration mechanism to maintain the higher accelerating field (~TV/m) at the rear side of target and consequently in achieving the maximum proton cutoff energy. These results indicate that peak proton energies >100 MeV can be achieved by limiting target extent and optimizing the laser beam focal spot with respect to the target thickness. High number of protons ($>10^9$) in the energy range 20 MeV < $E$ < 100 MeV have been observed which may be sufficient for many applications [7, 37-43].

The influence of target geometry beside the target dimensions, where it is demonstrated the ion energy dependence on the target shape (planar/cylindrical/spherical). The maximum proton energy for the rounded target (cylindrical and spherical) is several tens of MeV greater than the planar target. In this case, the laser pulse arrives on the target on larger angles giving rise to a more efficient collisionless absorption and to higher electron energies. In comparison with previous investigation [25] with the hydrogen gas target, an order of magnitude higher a conversion efficiency was obtained by employing the cryogenic hydrogen target. We also shown the influence of laser polarisation on proton beam characteristics, as a function of proton energy. It is also delineated that diffracted laser field beside the MLT target can shape the proton beam to make it appropriate for medical applications.

ACKNOWLEDGEMENTS

PIC simulations were performed using the open-source code PIConGPU version 0.1.2 [47]. We acknowledge support of the Department of Information Services and Computing, Helmholtz-Zentrum Dresden-Rossendorf (HZDR), Germany; for providing access to the GPU Compute Cluster Hypnos. The authors would also like to thank M. Bussmann and the PIConGPU developer team for fruitful discussions regarding the simulation work. PIConGPU is developed and maintained by the Computational Radiation Physics Group at the Institute of Radiation Physics, HZDR.

This work was funded as part of the European Cluster of Advanced Laser Light Sources (EUCALL) project which has received funding from the European Union's Horizon 2020 research and innovation programme under grant agreement No. 654220.